# Directed Transport of Atoms in a Hamiltonian Quantum Ratchet

**Tobias Salger**[1*], **Sebastian Kling**[1], **Tim Hecking**[1]
**Carsten Geckeler**[1], **Luis Morales-Molina**[2], **and Martin Weitz**[1]

[1]Institut für Angewandte Physik, Wegelerstr. 8, 53115 Bonn, Germany
[2]Facultad de Física, Pontificia Universidad Católica de Chile, Casilla 306, Santiago, Chile

[*]To whom correspondence should be addressed; E-mail: salger@iap.uni-bonn.de

**Abstract**

We demonstrate the operation of a quantum ratchet in the absence of dissipative processes within the observation time (Hamiltonian regime). An atomic rubidium Bose-Einstein condensate is exposed to a sawtooth-like optical lattice potential, whose amplitude is periodically modulated in time. The ratchet transport arises from broken spatiotemporal symmetries of the driven potential, resulting in a desymmetrisation of transporting Eigenstates (Floquet states). The measured atomic current oscillates around a non-zero stationary value at longer observation times, shows resonances at positions determined by the photon recoil and depends on the initial phase of the drive, providing different lines of evidence for the full quantum character of the ratchet transport. The results provide a proof of principle demonstration of a quantum motor.

Transport phenomena are at the heart of many problems in physics, chemistry and biology. An intriguing example is the ratchet effect, where in the absence of a biased force a directed current of particles is produced along a periodic structure (1). Much interest in ratchet systems has originated in their role as a proposed model for the operational principle of biological motors, tiny biological engines which transform the energy produced in chemical reactions into unidirectional mechanical motion (2). More generally, ratchets also provide models for microscopic machinery, allowing for a directed current in the presence of thermal noise far from thermal equilibrium (3). Experimentally, ratchet transport has been studied in solid state systems where some quantum effects were observed (4,5), and more recently in atomic physics systems (6-9). The system of cold atoms in optical lattices allows for studies of quantum transport in the effective absence of dissipation, when optical fields detuned far from resonance are used. In remarkable experiments, tunnelling oscillations and quantum resonances have been observed with moving lattice potentials (10-12). However, directed transport of particles that initially are at rest with respect to the lattice requires that the driving potential breaks certain symmetries (13). In recent ratchet experiments with laser-cooled atoms, the relationship between the appearance of transport and symmetry breaking has been investigated, and the scaling of transport versus dissipation was studied (6-9). Theoretical work has pointed out that ratchet transport should be possible also in Hamiltonian quantum systems, where no dissipative processes aid to break the time-inversion symmetry and thus ensure unidirectional atom transport (14-16). The interest in such proposed devices is that they would provide a blueprint for new quantum machinery (17).

We here report on evidence for atom transport in a Hamiltonian (non-dissipative) quantum ratchet. A rubidium Bose-Einstein condensate is exposed to a sawtooth-like optical lattice potential that is periodically modulated in time. When the driving lattice potential breaks spatial and temporal symmetries, directed transport of atoms is observed.

It is long known that in a flashing sawtooth potential as shown in Fig.1A ratchet transport is possible for a Brownian classical particle, for which directed transport proceeds by particles rolling down a potential hill, accumulating in a potential minimum and during the subsequent off-phase of the potential, thermal motion allows for transport from the location of one minimum to the next one (1). In contrast to ratchets based on dissipative forces and thermal motion, in a Hamiltonian quantum system relaxation processes cannot be used to ensure directed motion. For a quantum system periodically modulated in time, the dynamics can be described in terms of Eigenfunctions of the Hamiltonian: $H(t) = p^2/2m + V(x,t)$ with $V(x,t) = V(x,t+T)$, where T denotes the temporal periodicity. These so-called Floquet states are the analogs of energy Eigenstates in time-periodic quantum systems (18). Each of these states carries a distinct average momentum, see Fig.1B for a representation of two Floquet states in our flashing ratchet system. In a simplified picture, one can illustrate the Floquet states of nonzero average momentum as transporting conveyor belts moving in different directions along the lattice axis, as depicted in Fig.1C. After loading of an atomic Bose-Einstein condensate into the lattice, the atomic wavefunction can be expressed as a coherent superposition of different Floquet Eigenstates, where the relative phases and amplitudes are determined by the overlap to the BEC wavefunction. Directly after the loading, the relative phases are such that the total momentum of the quantum state equals that of the BEC, i.e. the atoms are still at rest. Subsequently, phase factors determining the wavefunction temporal evolution become important and we expect oscillatory motion due to the beating between Eigenstates. The corresponding Eigenfrequencies are incommensurate, and for larger times the atomic momentum oscillates around an average value determined by the weighted superposition of the momenta of the populated Eigenstates. Thus, an asymptotic directed transport can only be obtained when the populated Eigenstates have a non-zero average momentum, which requires a breaking of spatiotemporal symmetries. Besides the Floquet oscillation around a non-zero average value, two further benchmarks of Hamiltonian quantum ratchets are a dependence on the initial time of the external driving and discrete resonances at positions determined by the photon recoil (16). In previous works studying atomic quantum ratchet transport, none of these features has been experimentally validated (9).

In our experiment, an atomic rubidium Bose-Einstein condensate is loaded into a flashing ratchet potential of the form

$$V(x,t) = V(x) \cdot A(t), \qquad (1)$$

with $V(x) = V_1 \cos(2kx) + V_2 \cos(4kx+\phi)$, where $V_1$ and $V_2$ denote the amplitudes of the spatial lattice harmonics with periodicities $\lambda/2$ and $\lambda/4$ respectively synthesizing the ratchet potential, $k=2\pi/\lambda$ the wave vector of the driving optical field, and $\phi$ is the relative phase between the lattice harmonics. The fundamental spatial frequency of periodicity $\lambda/2$ is generated by a usual standing wave formed with two counterpropagating fields, where $\lambda$ denotes the wavelength of the driving laser, 1.5 nm ($10^5$ natural linewidths) detuned to the red of the D2-line. A lattice with spatial periodicity $\lambda/4$ is realized with the dispersion of Doppler-sensitive four-photon Raman-transitions between Zeeman ground state sublevels (19-21), driven by the same laser. The temporal modulation of the potential is described with a biharmonic function $A(t) = A_1\sin^2(\omega(t-t_0)/2) + A_2\sin^2(\omega(t-t_0)+\theta/2)$, where $A_1$ and $A_2$ denote the amplitudes of the harmonics respectively, $\omega = 2\pi/T$ the modulation frequency, $\theta$ the relative phase between harmonics and $t_0$ the initial time of the modulation, with $A(t<t_0)=0$. As shown in (13), a net transport in the flashing ratchet system requires one to break certain temporal and spatial symmetries (see Supporting Material). These conditions are fulfilled if $A_1, A_2 \neq 0$ and $\theta \neq 0, \pm\pi$ along with $V_1, V_2 \neq 0$ and $\phi \neq 0, \pm\pi$. After the interaction with the flashing ratchet potential, the atoms freely expand for a 15 ms period, after which an absorption image is recorded. By this time of flight technique, the atomic velocity distribution can be analyzed.

The interaction of atoms with the spatially periodic ratchet potential results in a diffraction into several discrete diffraction peaks. Typical experimental data are depicted in Fig.2A for an interaction time of 26 modulation periods and a temporal phase $\theta = -\pi/2$ (left), $\theta = 0$ (middle) and $\theta = \pi/2$ (right) between the drive frequency harmonics. The presence of a diffraction pattern with well separated orders ($\Delta p = 2\hbar k$ between adjacent peaks) shows that the atoms contributing to the peaks have not been subject to dissipation from photon scattering, which would smear out the diffraction peaks by at least a photon recoil. Both phases $\theta = -\pi/2$ and $\theta = \pi/2$ correspond to a temporally asymmetric driving of the ratchet, and the corresponding atomic diffraction patterns indeed display a small asymmetry of the diffraction pattern, mainly visible in the intensity of the $\pm 2$-nd diffraction orders.

For a more detailed analysis of the transport, the mean atomic momentum along the lattice axis is calculated from such time of flight images. Corresponding results are shown in Fig.2B as a function of the interaction time with the flashing ratchet potential. One clearly observes that the condensate, prepared initially at rest with respect to the lattice reference frame, is accelerated to positive (negative) momentum values for a phase between the two driving harmonics of $\theta = \pi/2$ ($\theta = -\pi/2$). The atoms continue to accelerate until a velocity of around 0.15 $\hbar$k/m (-0.15 $\hbar$k/m for opposite temporal phase) is reached after 30 periods of driving. For larger interaction times, an oscillation around a nonzero mean value is observed. The oscillation is attributed to the interference between discrete Floquet Eigenstates, which is in agreement with theoretical predictions for Hamiltonian quantum ratchet transport (16).

Results for the dependence of the atom transport on the spatial symmetry of the ratchet are displayed in Fig.3, showing the mean momentum as a function of the spatial phase between lattice harmonics for different values of the temporal phase. As expected, the directed transport is maximized when both phases $\phi$ and $\theta$ are near $\pm\pi/2$. The breaking of spatiotemporal symmetries is then largest, and the Floquet eigenstates are desymmetrized. In contrast, when $\phi = 0$ or $\theta = 0$, no clear directed transport is observed, which shows that no significant drifts are present in our experimental apparatus.

An important prediction for Hamiltonian quantum ratchets is a dependence of the current on the initial time $t_0$ (16). A variation in $t_0$ corresponds to a phase shift within the driving cycle. Fig.4A depicts experimental data for the atom transport as a function of this initial time, and the insets indicate the form of the modulation for two values of $t_0$. For a proper choice of the initial time an enhancement of the directed transport by some 30% can be achieved. Note that because of ergodicity in the chaotic layer of phase space, there is no dependence of the transport on the initial time expected in classical Hamiltonian or any dissipative ratchet systems (16). In contrast, Hamiltonian quantum systems memorize the initial condition.

For a further test of the quantum nature of the effect, the dependence of the ratchet transport on the modulation frequency $\omega$ of the flashing ratchet was studied. A tuning of the modulation frequency allows for critical control of the position of the (Floquet) Eigenstates of the system, and maximum transport is expected for strong mixing between such states, corresponding to a quantum resonance (1). Fig.4B shows measured data for the directed ratchet transport versus the drive frequency (see the black dots first). One observes a strong principle resonance near $\omega = 8\omega_r$ with $\omega_r = \hbar k^2/2m$ as the photon recoil frequency (at which the previously shown data sets were recorded) and a second resonance with oppositely directed transport at twice this value. In a simplified atom optics picture, the resonances can be understood by interpreting the flashing ratchet potential as a sequence of (temporally overlapping) optical beamsplitter pulses each diffracting atom waves, realizing a nested multiple path atom interferometer. The atoms interfere in several families of wavepackets, and a required condition for the different families to give a joint interference signal is that the path difference between adjacent paths reaches an integer multiple of $2\pi$, from which due to the presence of recoil phases quadratic in atomic momentum the same resonance condition can be derived (22).

To test whether the width of the atomic velocity distribution affects the linewidth of the observed resonances, we have in a subsequent measurement recorded data with reduced width of the initial atomic velocity distribution. The atomic Doppler width for a two-photon transition after release of the interacting condensate atoms from the trap equals 10 kHz, which can be reduced to 3 kHz by Bragg-selection of atoms followed by a Raman transfer pulse. Transport data recorded with the velocity selected atom source are shown by the red triangles in Fig. 4b versus drive frequency, which yields resonances with a clearly narrower spectral width. The velocity selection also increases the magnitude of the mean momentum of the directed transport to a value of 0.6 $\hbar k/m$, which is close to the expected results (16). We attribute the increased transport to the velocity width of the Bragg-selected atom source fitting better into the narrow velocity acceptance profile of the (Doppler-sensitive) quantum ratchet than that of the condensate atoms released directly from the trap.

To conclude, evidence for atom transport in a Hamiltonian quantum ratchet is obtained from (i) a saturation of the directed transport along with oscillations around the asymptotic current for an ensemble initially at rest, (ii) a dependence of the transport on the initial time of modulation and (iii) discrete resonances in the modulation frequency near positions determined by the photon recoil, as predicted in (16). The results may be interpreted as a quantum motor that provides directed transport of atoms in the absence of a net directed force, using operational principles fully quantum in nature. The results pave the way to a quantum motor delivering mechanical energy, as a further step towards quantum machinery (17).

**Supporting online Material**
**Materials and Methods**

A rubidium ($^{87}$Rb) Bose-Einstein condensate is produced all-optically by evaporative cooling in a quasistatic $CO_2$-laser dipole trap. During the final stages of the evaporation, a magnetic field gradient is activated, resulting in a spin-polarized condensate with typically $6 \cdot 10^4$ atoms in the $m_F = -1$ Zeeman component of the F = 1 hyperfine ground state.

Our method for generating a ratchet potential for cold atoms is as follows (19). For the fundamental spatial frequency, a conventional standing wave lattice potential is used, as achieved with two counterpropagating optical waves detuned from an atomic resonance. The resulting potential has the well-known spatial periodicity of $\lambda/2$, where $\lambda$ denotes the optical wavelength. In a quantum picture, the absorption of one photon of a running wave mode followed by the stimulated emission of a photon into a counterpropagating mode contribute to the trapping potential. A lattice with spatial periodicity of $\lambda/4$ could in principle be achieved by replacing each of the absorption and emission cycles with a stimulated process induced by two photons, as indicated in the right of Fig. S1A. Our scheme used for generation of the subwavelength lattice with periodicity of $\lambda/4$ is shown in Fig. S1B (19,21), that is based on a three-level atom. In this improved approach, absorption (stimulated emission) processes have been exchanged by stimulated emission (absorption) processes of an oppositely directed photon. The high resolution of Raman spectroscopy between two stable ground state sublevels $|g_-\rangle$ and $|g_+\rangle$ over an excited state $|e\rangle$ allows to clearly separate in frequency space the desired four-photon process from lower order contributions. The three-level atoms are irradiated by two optical beams of frequencies $\omega+\Delta\omega$ and $\omega-\Delta\omega$ from the left and by a beam of frequency $\omega$ from the right, see Fig S1B.

We use the rubidium $F = 1$ ground state Zeeman components $m_F = -1$ and 0 as levels $|g_-\rangle$ and $|g_+\rangle$, and the $5P_{3/2}$ manifold as the excited state $|e\rangle$. A magnetic bias field of 1.8 G removes the degeneracy of the Zeeman components. By combining lattice potentials of $\lambda/2$ and $\lambda/4$ spatial periodicities, a potential with variable spatial symmetry can be Fourier-synthesized. The light to generate the lattice potential is delivered by a tapered diode laser system. The emitted near-infrared beam is split into two, from which the two counterpropagating beams for generation of the periodic atom potentials are derived. In each of the beams, an acoustooptic modulator generates all required optical frequencies in the corresponding beam path. Moreover, we can control the potential depths of the two spatial harmonics individually and apply the desired temporal modulation of the ratchet potential according to eq. 1.

Following the notations of Ref. 13, there are four transformations that leave the Hamiltonian of the quantum ratchet system invariant. For the temporal case, if $A(t)$ is invariant under the transformation $A(t) = -A(t+T/2)$ (period $T=2\pi/\omega$), we can define a shift transformations (i) $S_a(x,p,t) \rightarrow (-x,-p,t+T/2)$. Moreover, if $A(t)$ fulfills the relation $A(t+\tau) = A(-t+\tau)$ for a time $\tau$, we can realize the transformation (ii) $S_b$: $(x,p,t) \rightarrow (x,-p,-t+2\tau)$. In order to break the required temporal symmetry, the two transformations have to be violated simultaneously, which is the case for $A_1, A_2 \neq 0$ and $\theta \neq 0, \pm\pi$. The spatial symmetry can be broken in the same way as described for the temporal case. The relative phases of the spatial and temporal harmonics of the driven potential are controlled by variation of the phases of the electronic function generators that provide the electronic drive frequencies of the acoustooptic modulators. The beams are sent through optical fibres and focused in a counterpropagating vertical geometry onto the atomic Bose-Einstein condensate.

After the preparation of the rubidium Bose-Einstein condensate, the $CO_2$-laser trapping beam is extinguished and the atoms are left in free ballistic fall. The vertically oriented lattice beams are activated 3 ms after the release of the atoms, from $CO_2$-laser trapping potential, which due to the lowering of the atomic density after expansion reduces the interatomic interactions. At this time, the interaction energy has been largely converted into kinetic energy, and the velocity width of the condensate reaches ±0.3 $\hbar k/m$, where m denotes the atomic mass. During the ballistic free fall of the atoms, the lattice beams induce a flashing ratchet potential to the atoms for a variable interaction time. The difference frequency between upwards and downwards propagating lattice beams is tuned acoustooptically to yield a spatially stationary lattice in a ballistically downwards falling frame. The lattice beams are then extinguished and the atomic momentum distribution is recorded with a time-of-flight absorption imaging technique.

For the data recorded with additional velocity selection, the atomic Bose-Einstein condensate was initially prepared in a magnetic field-insensitive ($m_F = 0$) Zeeman sublevel of the F=1 hyperfine ground state. We then applied a 350 μs long Bragg pulse to select a narrow velocity slice (±0.08 $\hbar k/m$ width) of the expanded condensate atoms, followed by a shorter (150 μs length) Doppler-free Raman pulse to transfer the selected atoms into the $m_F = 1$ Zeeman sublevel. The data with additional velocity selection was limited to a maximum of 60 cycles of the periodic ratchet drive, so that we cannot exploit the full region of ratchet interaction times investigated in Fig. 2B with the velocity selected atom source. For longer durations of the periodic drive, by the time of the optical detection pulse the atoms remaining in the $m_F = 0$ Zeeman ground state at available values of our pulsed Stern-Gerlach magnetic field used for state analysis spatially overlapped with atoms experiencing ratchet transport.

The quoted values for the potential depths of the two lattice harmonics have been measured by a series of measurements investigating Rabi oscillations (23). The amplitude and phase values for the temporal modulation of the driven ratchet potential were determined from the electronic drive signals of the acoustooptic modulators.

**Figure captions:**

Fig. 1 (A) Operation of a Brownian ratchet with classical particles. (B) Quasiclassical Husimi representation of the phase space for two different Floquet Eigenstates (top, bottom) of a Hamiltonian quantum ratchet (eq. 1). The used parameters are $\omega/8\omega_r = 1.05$, $A_1 = 0.78$, $A_2 = 0.27$, $V_1 = 6E_r$, $V_2 = 1.3E_r$ and $\phi \cong \pi/2$, where $E_r = \hbar\omega_r$ is the recoil energy, for $\theta \cong 0$ (left) and $\theta \cong \pi/2$ (right). In the latter case, a desymmetrisation of Floquet states is achieved. (C) Graphical scheme illustrating the loading of an atomic Bose-Einstein condensate onto Floquet Eigenstates of the flashing ratchet potential, visualized here as conveyor belts to represent the associated atom transport.

Fig. 2 (A) Time of flight images showing the atomic velocity distribution after 26 modulation periods of the ratchet potential for a relative phase between temporal harmonics of $\theta = -\pi/2$ (left), $\theta = 0$ (middle), and $\theta = \pi/2$ (right). The modulation period was $T = 2\pi/\omega \cong 32.1$ µs, for which $\omega \cong 8\omega_r$. Other parameters of the driving potential (eq. 1) were $V_1 = 6E_r$, $V_2 = 1.6E_r$, a phase $\phi = \pi/2$ and amplitudes of temporal harmonics $A_1 = 0.75$ and $A_2 = 0.40$. These values were chosen to maximize the directed transport. (B) Mean atomic momentum as a function of the number of modulation periods for $\theta = -\pi/2$ (blue triangles), $\theta = 0$ (black squares) and $\theta = \pi/2$ (red dots). The data points are fitted with a spline function.

Fig. 3 Mean atomic momentum versus the phase $\phi$ between spatial lattice harmonics for three different values of the phase between temporal harmonics of the flashing ratchet: $\theta = -\pi/2$ (blue triangles), $\theta = 0$ (black squares) and $\theta = \pi/2$ (red dots). To guide the eye, the data points have been fitted with sinusoidal curves. Parameters are the same as in Fig. 2A.

Fig. 4 (A) Mean atomic momentum as a function of the initial time of modulation $t_0$ of the flashing ratchet. (B) Mean atomic momentum as a function of the modulation frequency $\omega$ (black dots). The red triangles give data recorded with additional Bragg-selection of a narrowed atomic velocity slice. The data points have been fitted with Gaussian functions to guide the eye.

Fig. S1: (A) Left: Second order processes in a common standing wave lattice, yielding a spatial periodicity of λ/2 of the trapping potential. Right: Four-photon processes contributing to a λ/4 spatial periodicity potential. (B) Improved scheme for generation of a lattice potential of λ/4 spatial periodicity, as used in this work. In contrast to the scheme indicated on the right hand side of (A), unwanted two-photon standing wave processes are suppressed.

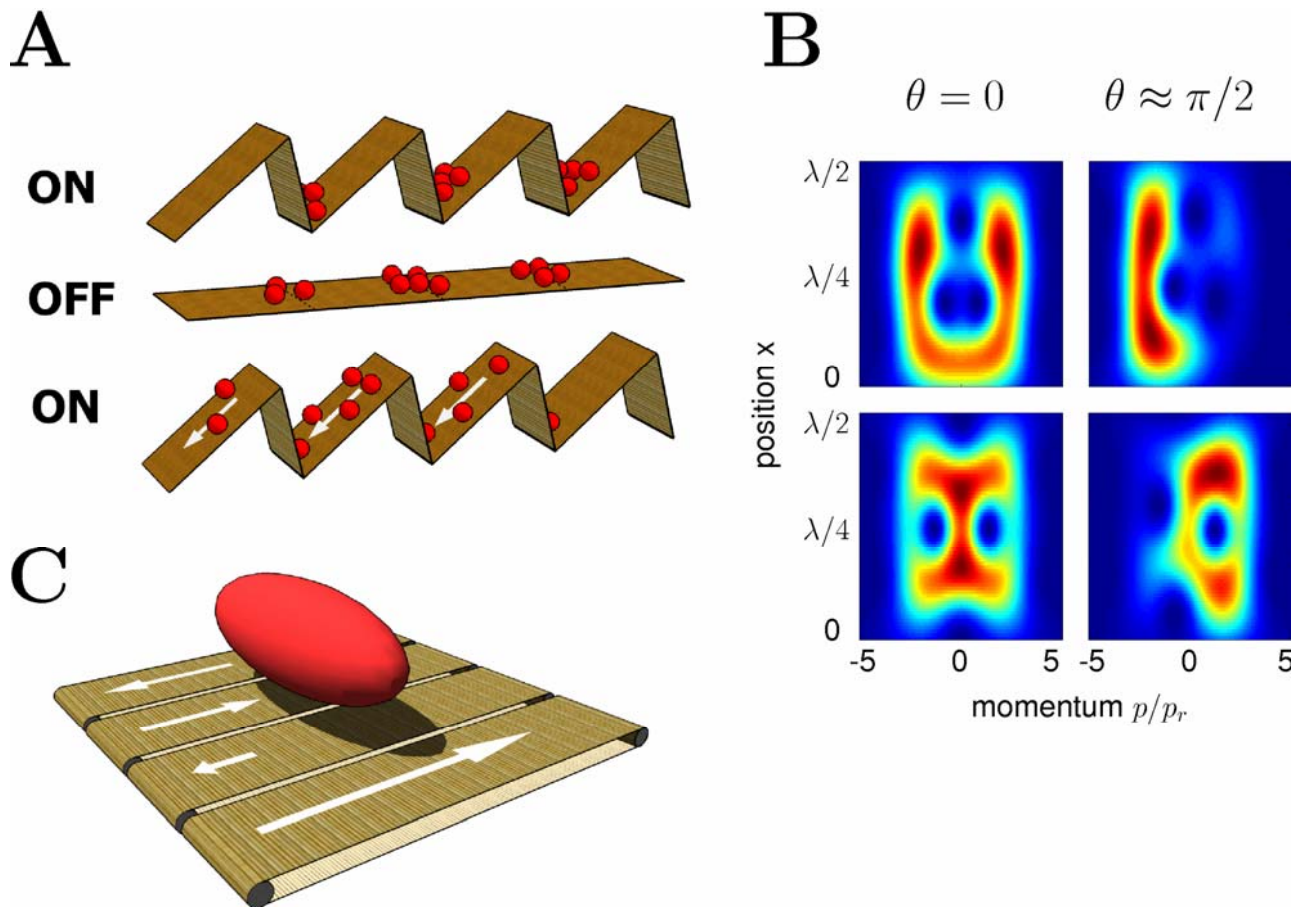

Figure 1

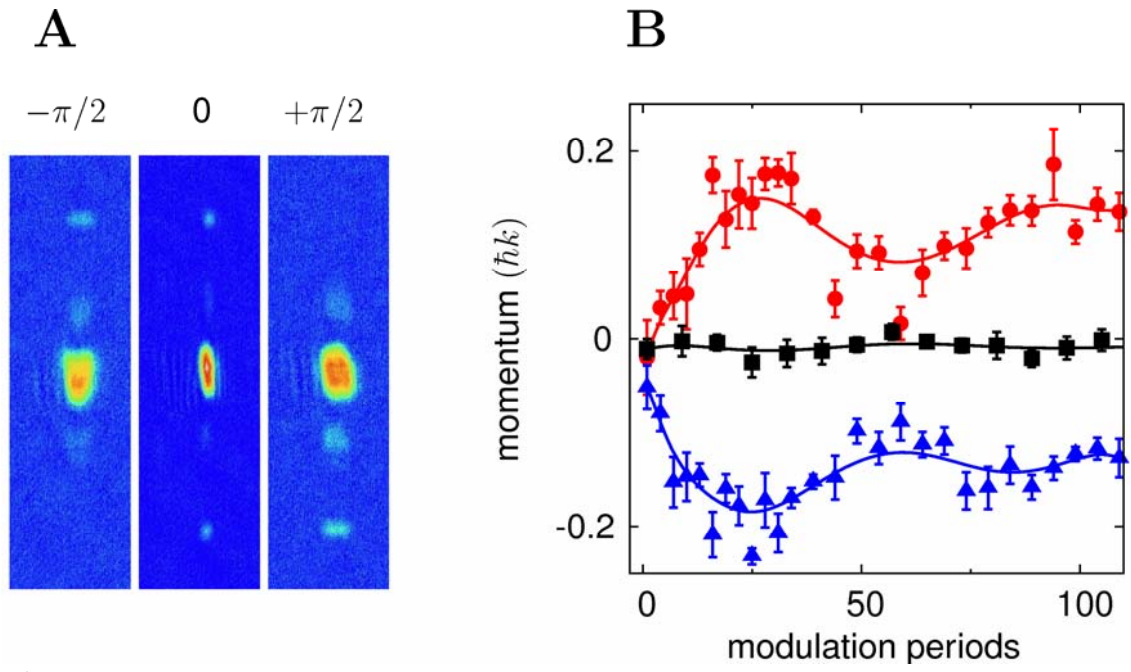

Figure 2

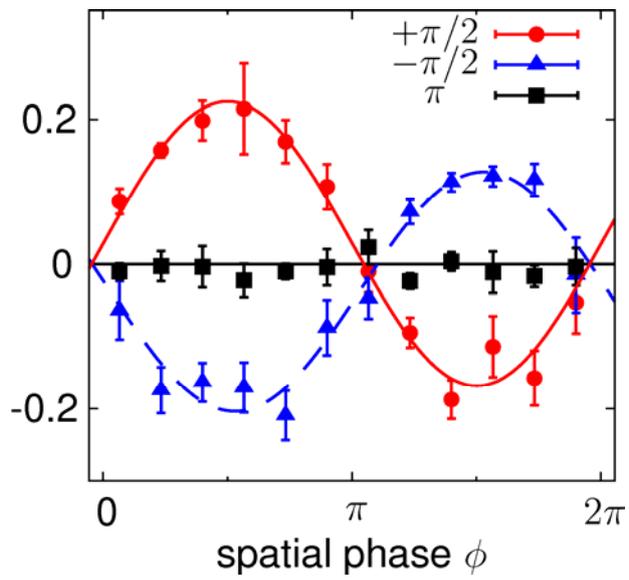

Figure 3

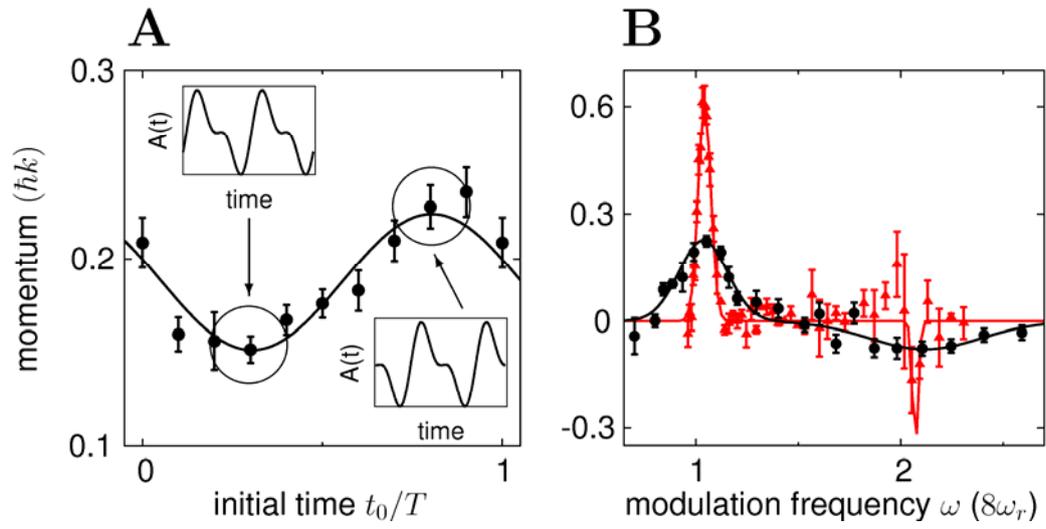

Figure 4

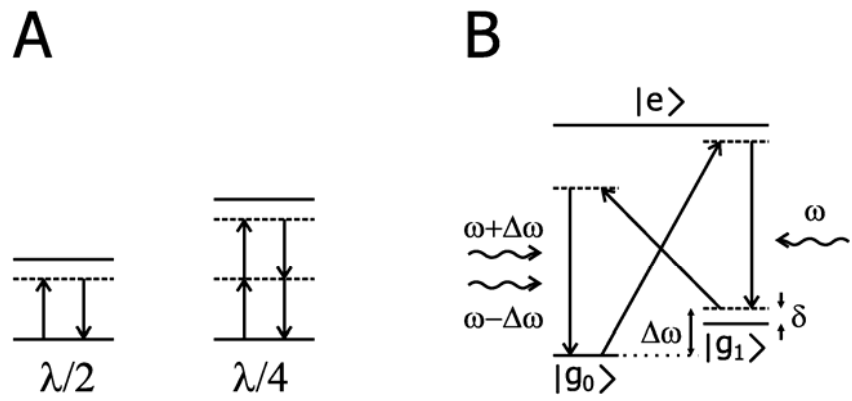

Supplementary Figure S1